\documentclass[journal]{IEEEtran}

\usepackage{cite}
\usepackage[T1]{fontenc}
\usepackage[latin9]{inputenc}
\usepackage{array}
\usepackage{float}
\usepackage{textcomp}
\usepackage{amsmath}
\usepackage{amssymb}
\usepackage{stackrel}
\usepackage{graphicx}
\usepackage{multirow}
\usepackage[normalem]{ulem}
\useunder{\uline}{\ul}{}
\usepackage[table,xcdraw]{xcolor}

\usepackage{lineno}
\usepackage{lipsum}
\modulolinenumbers[5]
\usepackage{url}

\usepackage{breakurl}
\usepackage[breaklinks]{hyperref}

\begin{document}

\title{Crowdfunding for Design Innovation: \\
Prediction Model with Critical Factors}

\author{Chaoyang~Song{*},~\IEEEmembership{Member,~IEEE,}
        Jianxi~Luo,
        Katja~H\"oltt\"a-Otto,
        Warren~Seering,
        and~Kevin~Otto,~\IEEEmembership{Member,~IEEE}
\thanks{{*}C. Song is the corresponding author with the Department of Mechanical and Energy Engineering, Southern University of Science and Technology, Shenzhen, China, 
    e-mail: \protect\href{mailto:songcy@ieee.org}{songcy@ieee.org}.}
\thanks{J. Luo is with Singapore University of Technology and Design.}
\thanks{K. H\"oltt\"a-Otto and K. Otto are with Aalto University.}
\thanks{W. Seering is with Massachusetts Institute of Technology.}
}

\markboth{A Preprint Accepted for IEEE Transactions on Engineering Management (DOI: 10.1109/TEM.2020.3001764)}%
{Song \MakeLowercase{\textit{et al.}}: Crowdfunding for Design Innovation}

\maketitle

\begin{abstract}
    Online reward-based crowdfunding campaigns have emerged as an innovative approach for validating demands, discovering early adopters, and seeking learning and feedback in the design processes of innovative products. However, crowdfunding campaigns for innovative products are faced with a high degree of uncertainty and suffer meager rates of success to fulfill their values for design. To guide designers and innovators for crowdfunding campaigns, this paper presents a data-driven methodology to build a prediction model with critical factors for crowdfunding success, based on public online crowdfunding campaign data. Specifically, the methodology filters 26 candidate factors in the Real-Win-Worth framework and identifies the critical ones via step-wise regression to predict the amount of crowdfunding. We demonstrate the methodology via deriving prediction models and identifying essential factors from 3D printer and smartwatch campaign data on Kickstarter and Indiegogo. The critical factors can guide campaign developments, and the prediction model may evaluate crowdfunding potential of innovations in contexts, to increase the chance of crowdfunding success of innovative products.
\end{abstract}

\begin{IEEEkeywords}
Crowdfunding, Design Innovation, Real-Win-Worth, Product Design, Entrepreneurship
\end{IEEEkeywords}

\IEEEpeerreviewmaketitle

\section{Introduction}
\label{S:1}
    \IEEEPARstart{M}{akers}, designers, innovators and entrepreneurs have increasingly adopted online crowdfunding campaigns to discover early users, validate design concepts, and collect design feedback for their innovative products as part of the design processes \cite{Massolution2015CrowdfundingPlatforms, Jensen2018IdentifyingProjects, Forbes2017GuidelinesCrowdfunding, Luo2018CrowdfundingEntrepreneurship}. Such benefits to design innovation are mainly provided by reward-based crowdfunding campaigns that engage early users of innovation via pre-ordering the first batch of products as rewards. In contrast, other types of crowdfunding are mainly useful for financing \cite{Sorenson2016ExpandCrowdfunding, Ahlers2015SignalingCrowdfunding, Agrawal2016AreCrowdfunding, Best2013HowEstimates}. Therefore, reward-based crowdfunding appears as an innovation in design processes and is the focus of this paper. Particularly, Kickstarter.com and Indiegogo.com have emerged as the most popular reward-based crowdfunding platforms, where many innovative product design projects raised a considerable amount of funding from the crowd via the Internet, such as Pebble on Kickstarter with \$ 10,266,845 raised \cite{PebbleTechnology2016Pebble:Kickstarter}, and Misfit Shine on Indiegogo with \$ 846,675 \cite{TheMisfitTeam2013MisfitIndiegogo}. 

    On online reward-based crowdfunding platforms, the creators publish their novel product concept and the state of product and project development through text, videos, figures, tables, as well as pledges for product delivery. If successfully raised, crowdfunding enables creators to continue product development and manufacturing, deliver the first batch of products to the crowd backers, and eventually enter the mass market. The ``funding'' is correlated with the engagement of early users and the learning that can be obtained via the interactions with them to inform the next design activities. On the other end, backers browse through campaign web pages and decide whether to make advanced payments in exchange for the promised products to be delivered soon (typically six months). Backers are mostly ``early adopters'' in the technology adoption life cycle \cite{Rogers1995DiffusionTelecommunications} and often provide useful feedback on novel design concepts. Crowdfunding platforms (CFPs) host campaigns, facilitate funding transactions, and provide communication channels between creators and backers via the Internet.

\begin{figure*}[h!]
    \centering
    \includegraphics[width=0.8\linewidth]{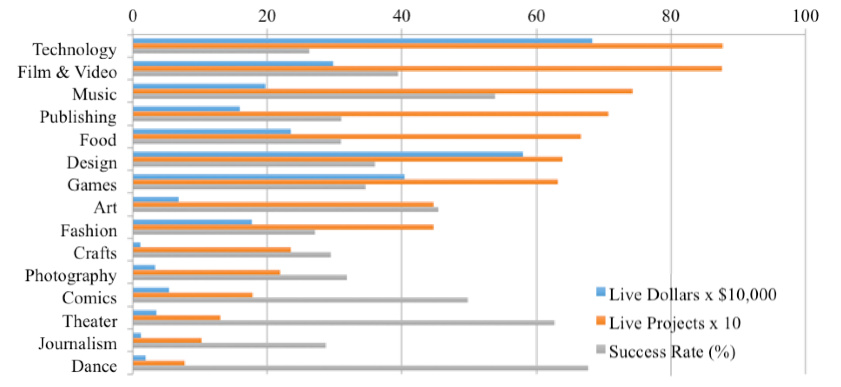}
    \caption{On Kickstarter, technology campaigns are presented with great interest for both creators and backers but suffer significantly from meager success rates. Only the top two ranked categories are labeled with numbers for each metric.}
    \label{fig:KickstarterStatistics}
\end{figure*}

    The innovation lies at the core of technology-based crowdfunding campaigns, which often comes at high risk. Most product concepts for online crowdfunding campaigns are characterized by a rather low level of technology maturity \cite{Belleflamme2014Crowdfunding:Crowd, Greenberg2013CrowdfundingFailure, Mollick2014TheStudy, Sharp2014}. The contradiction between innovation and risk is particularly evident in statistics from Kickstarter.com (Figure \ref{fig:KickstarterStatistics}). Campaigns in the technology category rank the highest among all fifteen categories in terms of total live projects (the creators' campaigns that are currently raising funding) and total live dollars (the backers' pledges to active campaigns). In other words, the technology category is the most popular for both creators and backers.

\begin{figure*}[h!]
    \centering
    \includegraphics[width=1.0\linewidth]{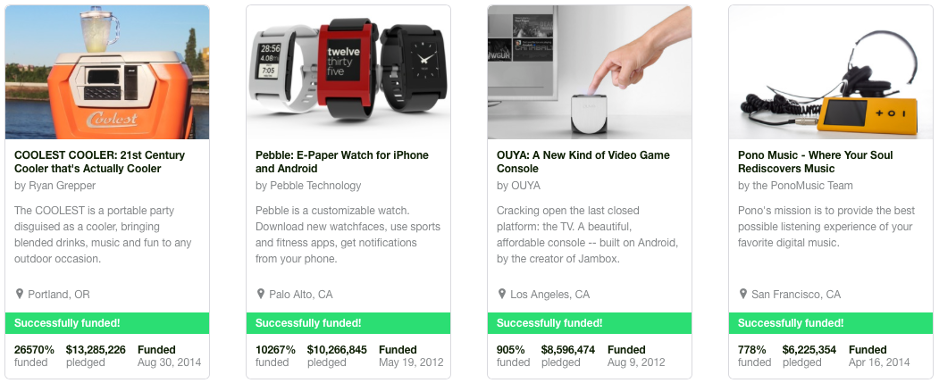}
    \caption{Screenshots of the top four most-funded campaigns on kickstarter.com.}
    \label{fig:KickstarterTopFunded}
\end{figure*}

    The most successful crowdfunding campaigns also appear to be technology-based. As illustrated in Figure \ref{fig:KickstarterTopFunded}, the top four most-funded campaigns on Kickstarter.com are all technology-based, most of which are consumer electronics. Unfortunately, technology campaigns also suffer the lowest ``success rate'' (the rate of campaigns that reach their funding goals) at 26.23\% among all categories on Kickstarter, as shown in Figure \ref{fig:KickstarterStatistics}. In other words, 74.76\% of the technology campaigns fail to obtain the requested development funding. The fact that technology campaigns are the most popular, with the most-funded star campaigns, but also have the lowest success rate suggests the need for methods or guidance to enhance such crowdfunding campaigns and ensure their values to design processes.

    To develop guidance for crowdfunding campaigns, we take a data-driven approach to derive a crowdfunding prediction model together with the critical product, team, or market factors for crowdfunding success, based on publicly available campaign data from online crowdfunding platforms. Our data-driven methodology is applied to identifying critical factors and training prediction models from 3D printer and smart watch campaign data on Kickstarter and Indiegogo. The prediction model estimates crowdfunding potentials of innovative products and the critical factors point to the core areas that require strategic attention and efforts in context. Both would make innovators more informed amid the high uncertainty of crowdfunding campaigns for their innovative products. Therefore, our research aims to contribute to the intersection of crowdfunding and design by focusing on innovation to study crowdfunding, and in turn, to guide crowdfunding campaigns to create value for design innovation.

    The remainder of this paper is organized as follows. In section \ref{S:2}, we first review how a crowdfunding campaign creates values for product innovation projects and specify the gaps in the prior research on crowdfunding. Section \ref{S:3} proposes and describes the methodology in detail. Section \ref{S:4} presents the case study. Section \ref{S:5} further discusses the methodological contributions, followed by the conclusion with limitations and future research in section \ref{S:6}.

\section{How Crowdfunding Works for Design Innovation}
\label{S:2}
    Through crowdfunding, creators raise relatively small amounts of money from many individuals through the Internet to fund the development of creative designs into innovative products \cite{Mollick2014TheStudy, tangibleengineeringUSACorporation2014SolidatorKickstarter}. According to what are in exchange for funding, there are four main types of crowdfunding: reward-based, equity-based, leading-based, and donation-based \cite{Belleflamme2014Crowdfunding:Crowd}. Based on a survey \cite{Massolution2015CrowdfundingPlatforms}, reward-based crowdfunding appears to be the most popular type. In addition to funding (in the form of advance payments of enthusiastic early users for initial products of a novel design concept), reward-based campaigns allow for demand testing of novel concepts, early adopter discovery and engagements, and need-finding and feedback generation, and thus are primarily relevant to design innovation. By contrast, other types of crowdfunding campaigns, such as donation-, leading- and equity-based crowdfunding, are irrelevant to the design processes and unable to inform design, other than financing \cite{Best2013HowEstimates, Sorenson2016ExpandCrowdfunding, Agrawal2016AreCrowdfunding}. Hereafter, we focus on reward-based crowdfunding campaigns for the interest of design processes and product innovation.

    Product innovation campaigns are usually found on reward-based crowdfunding platforms online, where the creators promise backers a unit from the first batch of products as rewards. Thus, the potential to deliver these rewards is critical for creators to attract crowdfunding backers \cite{Sharp2014, Zouhali-Worrall2011ComparisonWebsites}. For instance, each of the four most-funded Kickstarter campaigns (shown in Figure \ref{fig:KickstarterTopFunded}) presented an innovative product concept with a promising level of development maturity; these convinced backers that the reward (i.e., the innovative product) would be delivered if enough funding support could be secured. Meanwhile, additional factors may exist to affect crowd backers' perceptions and decisions. 

    On the other hand, backers play multiple roles in crowdfunding campaigns. For reward-based crowdfunding, backers collectively provide development and manufacturing funding by making advanced payments to pre-order a novel product, which usually does not interest regular financial investors enough to commit money and does not interest mainstream consumers enough because of the unfamiliarity of the new design. Crowdfunding campaign backers naturally are early adopters of innovative products. Such backers also often ask questions, share comments, and provide user feedback on the campaign web page. This feedback is highly valuable for creators to find design problems and user needs, and thus inspires or informs the creators about their next design tasks, opportunities, and directions. That is, reward-based crowdfunding campaigns provide an innovative channel for innovators to develop empathy toward users via Internet \cite{Brown2008DesignThinking, Luo2018CrowdfundingEntrepreneurship}. In contrast, traditional user research or empathy techniques are slow, expensive, and limited in terms of the scope and scale of the users that can be engaged.

    The story of Pebble's smart watch project demonstrates how crowdfunding works for product innovation. Its founder, Eric Migicovsky, initiated a one-person project in 2009 to develop an email notification device for Blackberry. Early development of this project received seed funding from Y Combinator and other angel investors \cite{Milian2012RejectedKickstarter}. In early 2012, Eric's team launched a reward-based crowdfunding campaign on Kickstarter. On its campaign web page, Pebble described the product's design concept as an e-paper smart watch and its wireless notification functionality with iOS and Android devices. The crowdfunding goal was set at \$ 100,000 within a one-month pledging period. Pebble raised \$ 10.3 million US dollars, making it the most-funded Kickstarter campaign at the time. During the campaign, Pebble received 15,629 comments from its backers as design feedback. With the demand validation and user feedback learning on Kickstarter, later Pebble attracted an additional \$ 15 million in Series A investment from venture capitalists \cite{Burns2013PebbleTechCrunch}. 

    Pebble was not the first to develop such a product. Large companies had launched similar products, such as Microsoft's Smart Personal Objects Technology (SPOT) watch between 2004 and 2008 \cite{Baker2010HowWrist-Top}. However, Pebble was the first to validate the market demands of such a product concept. Notably, this user demand validation was achieved via an online crowdfunding platform. Inspired by Pebble's success on Kickstarter, a series of smart watch products were launched later by not only startups but also incumbent firms such as Samsung, Google, and Apple. The crowdfunding campaigns of such large companies, which have abundant capitals, were not for funding but demand validation of new products with novel design concepts and user engagements for feedback. 

    In brief, the Pebble story shows that crowdfunding campaign via the Internet is an innovative and viable means for designers and innovators to validate market interest, discover early adopters and collect feedback for their new products faced with high uncertainty, in addition to funding. Therefore, a reward-based crowdfunding campaign is an innovative empathy technique for design thinking \cite{Brown2008DesignThinking}. Crowdfunding seamlessly synthesizes design thinking and financing for innovation \cite{Luo2018CrowdfundingEntrepreneurship}. In this manner, it also fulfills the ``lean startup'' strategy \cite{Ries2011TheBusinesses} by discovering and engaging a crowd of ``paying users'' for validated learning using minimal resources and does so using the Internet and online CFPs.

    However, in practice, only 26\% of the campaigns in the technology category on Kickstarter (see Figure \ref{fig:KickstarterStatistics}) achieve their funding goals, and even fewer achieve the level of success of Pebble or the others in Figure \ref{fig:KickstarterTopFunded}. If a campaign fails to attract many pre-orders and engage many backers, it will not be able to generate the promised values to design innovation. Therefore, creators would benefit by knowing what and how to make their campaigns effective to attract many backers and reach high funding levels. However, such guidance has not been seen in the literature.

    The rapid growth of crowdfunding has boosted academic research and related legislation such as the Jumpstart Our Business Startups Act (JOBS) \cite{USCongress2012JumpstartAct}. Primary research interests include the descriptive study of the crowdfunding phenomenon \cite{Forbes2017GuidelinesCrowdfunding, Agrawal2014}, the taxonomy of crowdfunding processes \cite{Belleflamme2014Crowdfunding:Crowd, Ahlers2015SignalingCrowdfunding}, policy facilitating crowdfunding \cite{Best2013HowEstimates, Zheng2014TheUS}, prediction of crowdfunding success \cite{Sharp2014} and sources of delays or cancellations of successfully crowdfunded product development \cite{Jensen2018IdentifyingProjects}, and applications of crowdfunding campaigns to entrepreneurship education \cite{Luo2018CrowdfundingEntrepreneurship}. Specifically, prediction-oriented studies have suggested that social networks \cite{Greenberg2013CrowdfundingFailure, Etter2013LaunchCampaigns} and campaign qualities \cite{Mollick2014TheStudy, Burtch2013AnMarkets} play essential roles in crowdfunding success. However, these studies have not differentiated technology campaigns from other types of campaigns and have not addressed the factors directly related to the innovative product and project itself. 

    Our research aims to fill these gaps by focusing on reward-based crowdfunding campaigns, analysis of the intrinsic characteristics of the innovation projects, and data-driven guidance for creators' campaign efforts. Therefore, this work takes the design innovation perspective, instead of a financial perspective, to study a crowdfunding campaign as an innovative part of the design innovation process.

\section{The Methodology}
\label{S:3}
    To deal with the uncertainty of crowdfunding campaigns for product innovation, we introduce a data-driven methodology to derive prediction models and identify critical factors for crowdfunding success, based on publicly available data of crowdfunding campaigns on the online crowdfunding platforms.

\subsection{Crowdfunding Data}
    Once published on online CFPs, the web pages of crowdfunding campaigns are permanently available regardless of the results of the campaigns. For each campaign, the creators provide a product and project description on the CFPs using text, figures, videos, and tables. The funding amount is also reported on each campaign's web page. As a result, rich and continually growing data can be found on both successful and failed crowdfunding campaigns on online CFPs. Such public data allow for training models to predict crowdfunding success based on the characteristics of the innovation projects. Herein, our prediction model is trained based on the intrinsic factors of the innovation projects that are critical and thus predictive for their crowdfunding success.

\subsection{Predictors of Crowdfunding Success}
    The studies of critical factors in new product development projects have a long tradition \cite{Evanschitzky2012SuccessMeta-Analysis} and have identified a comprehensive set of NPD success factors. Those factors can be leveraged as candidate critical factors in crowdfunding campaigns of innovative products, which are mostly new product development projects themselves. For instance, Cooper and his colleagues \cite{Evanschitzky2012SuccessMeta-Analysis, Cooper1993MajorIndustry, Cooper1987NewLosers, Cooper1979TheFailure} have identified various NPD success factors, including product advantage and uniqueness, market attractiveness, and internal organization. Montoya-Weiss and Calantone \cite{Montoya-Weiss1994DeterminantsMeta-analysis} suggested 18 success factors, with the main factors being product advantage and market synergy. Henard and Szymanski \cite{Henard2001WhyOthers} summarized 24 predictive factors for NPD success from the empirical research literature. They suggested factors that are more significant than others, including product characteristics (product advantage, product meeting customer needs, product technological sophistication), firm strategy (order of entry, dedicated human resources, dedicated R\&D resources), firm process (predevelopment task proficiency, marketing task proficiency, technological and launch proficiency), and marketplace characteristics (market potential). Ulrich, Eppinger and Yang \cite{Ulrich2019ProductDevelopment} suggested the general dimensions of factors that affect NPD performance in product cost, quality, development time, and capability.

    In this research, we adopt the ``Real-Win-Worth'' (RWW) framework \cite{Day2007} to screen and identify the critical factors that can best predict crowdfunding success, because it provides the most comprehensive coverage and systematic synthesis of the previously reported critical factors in the NPD literature. The RWW framework had been used by many established companies such as 3M and General Electric to evaluate internal innovation projects for go/kill decisions \cite{Day2007} and later modified to evaluate technology startups for accelerator selection \cite{Yin2018HowStages}. In particular, the RWW framework allows one to evaluate a wide spectrum of product, market, team, risk, and strategic factors of an innovation project by answering guiding questions in three main aspects \cite{Jensen2018IdentifyingProjects}:

\begin{itemize}
    \item ``\textbf{Is it Real?}'' evaluates market attractiveness and product feasibility;
    \item ``\textbf{Can We Win?}'' considers product advantage and team competency; 
    \item ``\textbf{Is it Worth Doing?}'' examines potential risk and strategic benefits.
\end{itemize}

    Six more specific queries address these central questions: Is the market real? Is the product real? Can the product be competitive? Can the team be competitive? Will the product be profitable at an acceptable risk? Does launching the product make strategic sense? To answer these six queries, one can explore an even more nuanced set of supporting questions. We developed 26 detailed questions to address 26 possible influential factors for crowdfunding success in the Real, Win and Worth categories and six subcategories, as shown in Table \ref{tab:RWWAdapted}. Answering these 26 questions based on campaign descriptions on a CFP leads to a systematic and structured evaluation of the innovation project from the angle of backers.

\begin{table*}[h!]
    \centering
    \caption{The adapted Real-Win-Worth metric with 26 questions for crowdfunding product campaigns.}
    \label{tab:RWWAdapted}
    \begin{tabular}{ccl}
    \hline
    \textbf{\begin{tabular}[c]{@{}c@{}}RWW Main\\ Categories\end{tabular}} &
      \textbf{\begin{tabular}[c]{@{}c@{}}RWW\\ Subcategories\end{tabular}} &
      \multicolumn{1}{c}{\textbf{\begin{tabular}[c]{@{}c@{}}Adapted \\ Questions\end{tabular}}} \\ \hline
    \multirow{11}{*}{Is it \textbf{Real}?} &
      \multirow{5}{*}{\begin{tabular}[c]{@{}c@{}}Market \\ Attractiveness\end{tabular}} &
      Q01. Is there adequate voice-of-customer type of evidence? \\ \cline{3-3} 
     &
       &
      Q02. Is there evidence of budget? \\ \cline{3-3} 
     &
       &
      Q03. Is there market demographic analysis evidence? \\ \cline{3-3} 
     &
       &
      Q04. Is there adequate evidence they understand the benefits? \\ \cline{3-3} 
     &
       &
      Q05. Is there adequate research on the subjective barriers that constrain the customer? \\ \cline{2-3} 
     &
      \multirow{6}{*}{\begin{tabular}[c]{@{}c@{}}Product \\ Feasibility\end{tabular}} &
      Q06. Is there evidence of adequate evolution of a product from an idea? \\ \cline{3-3} 
     &
       &
      \begin{tabular}[c]{@{}l@{}}Q07. Is there evidence of compatibility with existing local environment, including \\ regulatory compliance, legal \& social acceptability, and existing sales distribution channels?\end{tabular} \\ \cline{3-3} 
     &
       &
      \begin{tabular}[c]{@{}l@{}}Q08. Is there adequate evidence of functional feasibility with available/breakthrough \\ technology/material?\end{tabular} \\ \cline{3-3} 
     &
       &
      \begin{tabular}[c]{@{}l@{}}Q09. Is there adequate evidence that it can be produced and delivered with cost-efficiency \\ and manufacturability?\end{tabular} \\ \cline{3-3} 
     &
       &
      Q10. Is there adequate clarification of trade-offs in performance, cost, etc.? \\ \cline{3-3} 
     &
       &
      Q11. Is there adequate validation of the final product with market research on competitor positions? \\ \hline
    \multirow{10}{*}{Can we \textbf{Win}?} &
      \multirow{6}{*}{\begin{tabular}[c]{@{}c@{}}Product \\ Advantage\end{tabular}} &
      Q12. Is there adequate tangible or intangible advantages offered to the customers? \\ \cline{3-3} 
     &
       &
      Q13. Is there evidence showing that these advantages are not easily available to the competitors? \\ \cline{3-3} 
     &
       &
      Q14. Is there adequate patent strategy for existing/circumvent patents? \\ \cline{3-3} 
     &
       &
      Q15. Is there adequate company talent resources/channels to maintain the patent strategy? \\ \cline{3-3} 
     &
       &
      Q16. Is there an adequate evaluation of the vulnerabilities of the product advantages? \\ \cline{3-3} 
     &
       &
      Q17. Is there adequate evaluation of measures to cope with competitors? \\ \cline{2-3} 
     &
      \multirow{4}{*}{\begin{tabular}[c]{@{}c@{}}Team \\ Competency\end{tabular}} &
      \begin{tabular}[c]{@{}l@{}}Q18. Is there evidence of adequate resources to enhance the customer's perception of product \\ value and surpass the competitors?\end{tabular} \\ \cline{3-3} 
     &
       &
      Q19. Is there adequate market experience in the project leadership team? \\ \cline{3-3} 
     &
       &
      Q20. Is there adequate product development skillset in the project leadership team? \\ \cline{3-3} 
     &
       &
      Q21. Is there an adequate mechanism to listen and respond? \\ \hline
    \multirow{5}{*}{\begin{tabular}[c]{@{}c@{}}Is it \textbf{Worth} \\ doing?\end{tabular}} &
      \multirow{3}{*}{\begin{tabular}[c]{@{}c@{}}Expected \\ Return\end{tabular}} &
      Q22. Is there evidence of adequate profitability? \\ \cline{3-3} 
     &
       &
      Q23. Is there evidence of adequate cash flow robustness to changes in market, price and timing? \\ \cline{3-3} 
     &
       &
      Q24. Is there evidence of adequate measures to mitigate the potential product failures? \\ \cline{2-3} 
     &
      \multirow{2}{*}{\begin{tabular}[c]{@{}c@{}}Strategic \\ Fit\end{tabular}} &
      Q25. Is there adequate evidence that the product supports an overall growth strategy? \\ \cline{3-3} 
     &
       &
      Q26. Is there evidence of adequate agreement in project assumptions? \\ \hline
    \end{tabular}
\end{table*}

    With the 26 guiding questions, one can read the campaign descriptions on the CFPs to decide whether evidence of the existence of each of these 26 RWW factors is Full, Partial, or None. Then, these Full, Partial, and None ratings are transformed into the scores of 1, 0.5, and 0, respectively, for statistical analysis. Such numeric ratings measure the evidence about the state of product and project development in the eyes of potential backers. In addition to the guiding questions, we also developed rating criteria to guide the raters and ensure answers' independence from raters.

    To this end, three researchers first independently evaluated and scored three sample campaigns against the 26 questions (in their initial version), and then intensively discussed and compared their rating criteria and ratings, as a process to co-develop the rating criteria for each of the 26 questions and also refine the questions. The guiding questions were fine-tuned, and rating criteria are developed to reconcile the raters' different interpretations of the RWW questions and the varied availability of empirical evidence on the CFPs for answering the RWW questions. Using the updated questions and synchronized rating criteria, the three researchers rated the same sample campaigns again, and inter-rater repeatability reached an acceptable level as indicated by Cohen's Kappa ratio \cite{Carletta1996AssessingStatistic}. Next, an additional researcher rated the same sample of three campaigns using the refined guiding questions and rating criteria agreed upon by the first three researchers. Comparing the ratings of the new rater with previous ones, a weighted Kappa ratio of 80\% was reached, indicating that the rating process using the fine-tuned 26 questions (Table \ref{tab:RWWAdapted}) and corresponding rating criteria (examples in Table \ref{tab:RWWExamples}) is reasonably repeatable and the results are rater independent.

\begin{table*}[h!]
    \centering
    \caption{Examples of the rating process using the RWW questions and rating criteria based on the crowdfunding campaign description online. Rating criteria for all questions are available upon request.}
    \label{tab:RWWExamples}
    \begin{tabular}{clll}
    \hline
    \textbf{\begin{tabular}[c]{@{}c@{}}RWW \\ Questions\end{tabular}} &
      \textbf{\begin{tabular}[c]{@{}l@{}}Q01. Is there adequate \\ voice-of-customer type \\ of evidence?\end{tabular}} &
      \textbf{\begin{tabular}[c]{@{}l@{}}Q12. Is there adequate tangible \\ or intangible advantages offered \\ to the customers?\end{tabular}} &
      \textbf{\begin{tabular}[c]{@{}l@{}}Q25. Is there adequate evidence \\ that this product makes strategic \\ sense?\end{tabular}} \\ \hline
    \multirow{3}{*}{\textbf{Rating Criteria}} &
      \begin{tabular}[c]{@{}l@{}}\textbf{Full}: Four or more customer \\ observations, interviews, surveys \\ (not counting self-observation)\end{tabular} &
      \begin{tabular}[c]{@{}l@{}}\textbf{Full}: Both convincing tangible benefits \\ (lifetime cost saving, safety, quality, \\ maintenance support), and convincing \\ intangible benefits (social acceptability, \\ known brand) offered\end{tabular} &
      \begin{tabular}[c]{@{}l@{}}\textbf{Full}: A long-term growth strategy \\ such as product roadmap, \\ considering stakeholder and \\ opportunities\end{tabular} \\ \cline{2-4} 
     &
      \begin{tabular}[c]{@{}l@{}}\textbf{Partial}: Experienced self-observation, \\ or 1$\sim$3 customer interviews, surveys, \\ performance gap data, counting \\ Facebook similar likes or others\end{tabular} &
      \begin{tabular}[c]{@{}l@{}}\textbf{Partial}: Either convincing tangible \\ benefits (lifetime cost saving, safety, \\ quality, maintenance support) or \\ convincing intangible benefits \\ (social acceptability, known brand) \\ offered\end{tabular} &
      \begin{tabular}[c]{@{}l@{}}\textbf{Partial}: A short-term growth \\ strategy such as a discussion \\ of future derivatives\end{tabular} \\ \cline{2-4} 
     &
      \begin{tabular}[c]{@{}l@{}}\textbf{None}: Not determined yet or \\ less than Partial\end{tabular} &
      \begin{tabular}[c]{@{}l@{}}\textbf{None}: Not mentioned in any form at all; \\ or the benefits described are not \\ considered as strong and convincing \\ enough\end{tabular} &
      \begin{tabular}[c]{@{}l@{}}\textbf{None}: No mention or further \\ details on future options to \\ consider\end{tabular} \\ \hline
    \begin{tabular}[c]{@{}c@{}}\textbf{Description of} \\ \textbf{''Form 1'' 3D} \\ \textbf{Printer on} \\ \textbf{Kickstarter}\\ \\ (Rating is based \\ on the bold \\ italic content)\end{tabular} &
      \begin{tabular}[c]{@{}l@{}}''Our reason for starting this \\ project is simple: there are \\ no low-cost 3D printers that \\ meet the quality standards of \\ the professional designer. \\ \textbf{As researchers at the MIT} \\ \textbf{Media Lab, we were lucky} \\ \textbf{to experience the best and} \\ \textbf{most expensive fabrication} \\ \textbf{equipment in the world. But,} \\ \textbf{we became frustrated by the} \\ \textbf{fact that all the professional-}\\ \textbf{quality 3D printers were} \\ \textbf{ridiculously expensive (read:} \\ \textbf{tens of thousands of dollars)} \\ \textbf{and were so complex to use.} \\ In 2011, we decided to build \\ a solution to this problem \\ ourselves, and we are now \\ ready to share it with the \\ world.''\end{tabular} &
      \begin{tabular}[c]{@{}l@{}}''We've gone to extraordinary lengths \\ to design \textbf{a complete 3D printing} \\ \textbf{experience:}\\ - The Form 1 printer is engineered \\ \textbf{to produce high resolution parts with} \\ \textbf{the touch of a button}\\ - Form software is \textbf{intuitive and} \\ \textbf{simple to use} so you can spend \textbf{less} \\ \textbf{time} setting up prints and \textbf{more time} \\ designing\\ - The Form Finish \textbf{post-processing} \\ \textbf{kit keeps your desktop organized} so \\ that you can easily put the finishing \\ touches on your masterpiece\\ Read on for \textbf{more details} on what \\ you'll help bring into the world (and \\ onto your desktop) if you support \\ this effort.''\end{tabular} &
      \begin{tabular}[c]{@{}l@{}}''\textbf{When} the Form 1 is \textbf{released}, \\ it will come with our \textbf{first material} \\ - a neutral matte gray that is great \\ for look-and-feel models, \\ standalone parts, or even as a base \\ color for painting.\\ \textbf{After a successful launch} (thanks \\ to your support!), we will \textbf{continue} \\ \textbf{development of an entire palette of} \\ \textbf{materials} for your printer. A variety \\ of colors, transparency, flexibility, \\ and even burnout capability for lost \\ wax casting processes are all \\ possible with SL.''\end{tabular} \\ \hline
    \textbf{Rating} &
      \multicolumn{1}{c}{Partial (0.5)} &
      \multicolumn{1}{c}{Full (1.0)} &
      \multicolumn{1}{c}{Partial (0.5)} \\ \hline
    \end{tabular}
\end{table*}

\subsection{Measure of Crowdfunding Success}
    Then the numeric ratings of 26 factors of innovation campaigns are used to predict their crowdfunding success. To measure the crowdfunding success of a campaign, the amount of funding raised in its natural log form is used. In contrast, prior studies often used a binary variable to denote whether the funding goal had been reached (1) or not (0). Focusing on the actual amount of funding offers two benefits. First, the amount of funding raised provides a more accurate measure of the backers' interest and the level of engagements with early adopters, compared to the binary variable. Campaigns that ``fail'' to reach the funding goals may raise more funding than other similar campaigns that ``succeed'' to meet the funding goals with less funding raised. Table \ref{tab:FundedExamples} presents such comparative cases. For example, Coolest Cooler and Ubuntu Edge are among the most funded campaigns on Kickstarter and Indiegogo, respectively. Coolest Cooler is considered a significant crowdfunding success on Kickstarter with 26,570\% funded percent. However, Ubuntu Edge was a failed campaign, despite its phenomenal amount of funding raised, only because its funding goal was set too high. Therefore, the binary measure of crowdfunding success versus failure is unable to reflect the actual level of interest from backers.

\begin{table*}[h!]
    \centering
    \caption{Exemplar crowdfunding campaigns with varied funding goals.}
    \label{tab:FundedExamples}
    \begin{tabular}{lllll}
    \hline
    \textbf{\begin{tabular}[c]{@{}l@{}}Campaign \\ Title\end{tabular}} &
      \textbf{\begin{tabular}[c]{@{}l@{}}Lowest \\ Unit Price\end{tabular}} &
      \textbf{\begin{tabular}[c]{@{}l@{}}Funding \\ Goal\end{tabular}} &
      \textbf{\begin{tabular}[c]{@{}l@{}}Funding \\ Raised\end{tabular}} &
      \textbf{\begin{tabular}[c]{@{}l@{}}Funded \\ Percent\end{tabular}} \\ \hline
    Potato Salad by Brown \cite{Brown2014PotatoKickstarter}                                                                                            & \$ 3     & \$ 10         & \$ 55,492     & 554,928\% \\ \hline
    \begin{tabular}[c]{@{}l@{}}COOLEST COOLER: 21st Century Cooler \\ that's Actually Cooler by Grepper \cite{Grepper2014COOLESTKickstarter} \end{tabular} & \$ 165   & \$ 50,000     & \$ 13,285,226 & 26,570\%  \\ \hline
    \begin{tabular}[c]{@{}l@{}}FORM 1: An affordable, professional 3D printer \\ by Formlabs \cite{tangibleengineeringUSACorporation2014SolidatorKickstarter} \end{tabular}            & \$ 2,299 & \$ 100,000    & \$ 2,945,885  & 2,946\%   \\ \hline
    \begin{tabular}[c]{@{}l@{}}Solidator DLP Desktop 3D Printer by tangible \\ engineering USA \cite{Formlabs2014FORMKickstarter} \end{tabular}          & \$ 4950  & \$ 125,000    & \$ 144,403    & 116\%     \\ \hline
    Ubuntu Edge by Canonical \cite{Canonical2013UbuntuIndiegogo}                                                                                         & \$ 695   & \$ 32,000,000 & \$ 12,814,216 & 40\%      \\ \hline
    \multicolumn{5}{l}{\begin{tabular}[c]{@{}l@{}}Note: The ''lowest unit price'' is the lowest price the creators offer for the product reward and excludes the \\ nonproduct rewards, such as a ''thank you'' note and stickers.\end{tabular}}
    \end{tabular}
\end{table*}

    Second, funding rules of different platforms are different and need to be reconciled in the dependent variable. For instance, Kickstarter uses an all-or-nothing funding rule, meaning that the creators will receive all funds raised as long the pre-set funding goal is reached, or they will get nothing. This is the same as the fixed funding rule on Indiegogo. However, Indiegogo also provides a flexible funding option so that the creators can collect any amount of funds raised, regardless of whether the funding goal is reached. Prior crowdfunding research primarily used data from only one source - Kickstarter \cite{Mollick2014TheStudy, Mollick2016DemocratizingCrowdfunding, Jensen2018IdentifyingProjects, Greenberg2013CrowdfundingFailure}. Therefore, to study different platforms, a generic measure of crowdfunding success is needed.

\subsection{Prediction Model}
    Then the RWW factors are incorporated as predictive variables in step-wise regressions to train a regression model that achieves the highest predictability on the crowdfunding amount. The step-wise regression procedure inserts the candidate predictive variables into or removes the variables from the trial regression model in a step-wise manner to fine-tune the model regarding the statistical fit, i.e., $R^2$. The most predictive regression model that results might include a subset, not all the candidate predictive variables.
    
    In the step-wise regressions with varied predictors, we control for the following factors exogenous to the product and innovation project itself, such as the description text length, the number of videos, figures, and tables. These exogenous factors had been found influential to crowdfunding according to prior studies \cite{Mollick2016DemocratizingCrowdfunding, Mollick2014TheStudy}.

\begin{itemize}
    \item \textit{The number of characters in campaign description}: This variable addresses the length of the description text for the crowdfunding campaign. The natural log of this variable is used. 
    \item \textit{The number of figures, videos, tables, and rewards}: The appearance of each described objects in the campaign description section of a crowdfunding campaign.
    \item \textit{Team introduction}: This binary variable is 1 (or 0, otherwise) if there is a description of the team members, their experiences, and responsibilities in the campaign description section. 
    \item \textit{Timeline}: This binary variable is 1 (or 0, otherwise) if there is a description of the campaign schedule, such as when to finish design, arrange production, and deliver the rewards, in the campaign description section.
\end{itemize}

    During the step-wise regression, although the candidate predictive variables (i.e., the 26 factors) were removed or added in a step-wise manner, the control variables above were always included in all intermediate regression models in the search for the best model. Such regression models use the critical factors (characterizing the innovation project itself) as well as the control variables (covering the influences of exogenous factors) to explain crowdfunding success.

\section{Case Study}
\label{S:4}
\subsection{Empirical Context and Data}
    We applied the methodology to the empirical contexts of 3D printer and smart watch campaigns on Kickstarter.com and Indiegogo.com to derive prediction models with critical factors for these crowdfunding contexts. Kickstarter and Indiegogo are two leading reward-based CFPs \cite{Zouhali-Worrall2011ComparisonWebsites}. Product designers and technology innovators normally choose these two CFPs to launch their reward-based campaigns for demand validation and design feedback. Key characteristics and differentiation of these two CFPs include: 1) Kickstarter promotes campaigns with creativity whereas Indiegogo encourages everyone to participate; 2) Kickstarter uses the funding rule of all-or-nothing, but Indiegogo provides both fixed and flexible funding options (see the next section); and 3) Kickstarter has been more successful than Indiegogo in the amount of funding raised, the number of projects enlisted, and the funding success rate \cite{Lau2013DollarOver}.
    
    3D printers and smart watches are two popular crowdfunding product types. The 3D printers found in the crowdfunding campaigns are all consumer-level electronics for additive manufacturing using technologies such as Fused Deposition Modeling (FDM), stereolithography (SLA) and selective laser sintering (SLS) to build three-dimensional objects by successively adding materials layer by layer. The smart watch is a wearable electronic product for personal usage, taking the form of a wristwatch. Besides telling time, it is usually connected wirelessly with a smartphone for notification, personal health monitoring, or more advanced functions such as making phone calls or instant messaging.
    
    Via an exhaustive search for relevant campaigns, we created a dataset of 127 campaigns, including 47 3D printers and 23 smart watches from Kickstarter, and 31 3D printers and 26 smart watches from Indiegogo. In our dataset, more than half of the Kickstarter campaigns reached their funding goals, while the opposite was true for the Indiegogo. More than half of the 3D printer campaigns reached their funding goals, with the opposite being true for the smart watch campaigns. With the dataset, our analysis contrasts to those prior studies that do not differentiate the product types and that focus on only one platform (normally Kickstarter) \cite{Mollick2014TheStudy, Mollick2016DemocratizingCrowdfunding, Greenberg2013CrowdfundingFailure} or only successfully funded campaigns excluding failed ones \cite{Jensen2018IdentifyingProjects}.

\begin{table*}[h!]
    \centering
    \caption{Average ratings based on the RWW framework (1=Full, 0.5=Partial, 0=None).}
    \label{tab:RWWAvgRating}
    \begin{tabular}{ccllllll}
    \hline
    \multicolumn{3}{c}{\textbf{RWW Factors}} &
      \multicolumn{1}{c}{\textbf{3D Printer}} &
      \multicolumn{1}{c}{\textbf{Smart Watch}} &
      \multicolumn{1}{c}{\textbf{Kickstarter}} &
      \multicolumn{1}{c}{\textbf{Indiegogo}} &
      \multicolumn{1}{c}{\textbf{All}} \\ \hline
    \multirow{11}{*}{Real} &
      \multirow{5}{*}{\begin{tabular}[c]{@{}c@{}}Market \\ Attractiveness\end{tabular}} &
      Q01-voice of customer &
      0.32 (0.04) &
      0.42 (0.03) &
      0.37 (0.03) &
      0.39 (0.04) &
      0.37 (0.02) \\ \cline{3-8} 
     &
       &
      Q02-budget analysis &
      0.02 (0.01) &
      0.00 (0.00) &
      0.00 (0.00) &
      0.02 (0.01) &
      0.01 (0.01) \\ \cline{3-8} 
     &
       &
      Q03-market demography &
      0.17 (0.03) &
      0.24 (0.03) &
      0.21 (0.03) &
      0.21 (0.04) &
      0.21 (0.02) \\ \cline{3-8} 
     &
       &
      Q04-benefits understood &
      0.46 (0.03) &
      0.68 (0.03) &
      0.59 (0.03) &
      0.56 (0.04) &
      0.58 (0.02) \\ \cline{3-8} 
     &
       &
      Q05-subjective barrier &
      0.37 (0.04) &
      0.53 (0.03) &
      0.41 (0.04) &
      0.53 (0.04) &
      0.46 (0.03) \\ \cline{2-8} 
     &
      \multirow{6}{*}{\begin{tabular}[c]{@{}c@{}}Product \\ Feasibility\end{tabular}} &
      Q06-concept evolution &
      0.50 (0.05) &
      0.79 (0.03) &
      0.61 (0.04) &
      0.73 (0.05) &
      0.66 (0.03) \\ \cline{3-8} 
     &
       &
      Q07-development compatibility &
      0.12 (0.03) &
      0.49 (0.03) &
      0.33 (0.03) &
      0.31 (0.04) &
      0.32 (0.03) \\ \cline{3-8} 
     &
       &
      Q08-functional feasibility &
      0.46 (0.06) &
      0.51 (0.06) &
      0.41 (0.05) &
      0.61 (0.07) &
      0.49 (0.04) \\ \cline{3-8} 
     &
       &
      Q09-cost-efficient manufacturing &
      0.28 (0.05) &
      0.60 (0.04) &
      0.46 (0.04) &
      0.46 (0.06) &
      0.46 (0.03) \\ \cline{3-8} 
     &
       &
      Q10-clarified tradeoffs &
      0.26 (0.04) &
      0.42 (0.03) &
      0.35 (0.03) &
      0.35 (0.05) &
      0.35 (0.03) \\ \cline{3-8} 
     &
       &
      Q11-competition validation &
      0.17 (0.04) &
      0.29 (0.03) &
      0.22 (0.03) &
      0.26 (0.04) &
      0.24 (0.02) \\ \hline
    \multirow{10}{*}{Win} &
      \multirow{6}{*}{\begin{tabular}[c]{@{}c@{}}Product \\ Advantage\end{tabular}} &
      Q12-value propositions &
      0.25 (0.04) &
      0.54 (0.04) &
      0.38 (0.04) &
      0.44 (0.05) &
      0.41 (0.03) \\ \cline{3-8} 
     &
       &
      Q13-unique advantage &
      0.08 (0.03) &
      0.07 (0.02) &
      0.06 (0.02) &
      0.09 (0.03) &
      0.07 (0.02) \\ \cline{3-8} 
     &
       &
      Q14-patent strategy &
      0.04 (0.02) &
      0.05 (0.02) &
      0.01 (0.01) &
      0.09 (0.03) &
      0.04 (0.01) \\ \cline{3-8} 
     &
       &
      Q15-patent maintenance &
      0.11 (0.04) &
      0.16 (0.03) &
      0.12 (0.03) &
      0.17 (0.04) &
      0.14 (0.02) \\ \cline{3-8} 
     &
       &
      Q16-risk evaluation &
      0.16 (0.03) &
      0.10 (0.02) &
      0.12 (0.02) &
      0.13 (0.04) &
      0.13 (0.02) \\ \cline{3-8} 
     &
       &
      Q17-competition measures &
      0.02 (0.01) &
      0.00 (0.00) &
      0.00 (0.00) &
      0.02 (0.01) &
      0.01 (0.01) \\ \cline{2-8} 
     &
      \multirow{4}{*}{\begin{tabular}[c]{@{}c@{}}Team \\ Competency\end{tabular}} &
      Q18-enhanced perception &
      0.09 (0.03) &
      0.21 (0.03) &
      0.15 (0.03) &
      0.16 (0.03) &
      0.15 (0.02) \\ \cline{3-8} 
     &
       &
      Q19-leadership in marketing &
      0.08 (0.03) &
      0.06 (0.02) &
      0.04 (0.02) &
      0.10 (0.04) &
      0.07 (0.02) \\ \cline{3-8} 
     &
       &
      Q20-leadership in PD &
      0.11 (0.03) &
      0.18 (0.04) &
      0.14 (0.03) &
      0.15 (0.04) &
      0.15 (0.03) \\ \cline{3-8} 
     &
       &
      Q21-feedback management &
      0.01 (0.01) &
      0.16 (0.03) &
      0.10 (0.03) &
      0.07 (0.03) &
      0.09 (0.02) \\ \hline
    \multirow{5}{*}{Worth} &
      \multirow{3}{*}{\begin{tabular}[c]{@{}c@{}}Expected \\ Return\end{tabular}} &
      Q22-understood profitability &
      0.00 (0.00) &
      0.00 (0.00) &
      0.00 (0.00) &
      0.00 (0.00) &
      0.00 (0.00) \\ \cline{3-8} 
     &
       &
      Q23-cash flow robustness &
      0.17 (0.03) &
      0.44 (0.02) &
      0.32 (0.03) &
      0.31 (0.04) &
      0.31 (0.02) \\ \cline{3-8} 
     &
       &
      Q24-failure migration &
      0.08 (0.02) &
      0.39 (0.03) &
      0.27 (0.03) &
      0.21 (0.04) &
      0.25 (0.02) \\ \cline{2-8} 
     &
      \multirow{2}{*}{\begin{tabular}[c]{@{}c@{}}Strategic \\ Fit\end{tabular}} &
      Q25-growth alignment &
      0.24 (0.04) &
      0.29 (0.03) &
      0.27 (0.03) &
      0.27 (0.05) &
      0.27 (0.03) \\ \cline{3-8} 
     &
       &
      Q26-agreed management &
      0.03 (0.02) &
      0.03 (0.01) &
      0.03 (0.02) &
      0.02 (0.01) &
      0.03 (0.01) \\ \hline
    \multicolumn{8}{l}{Note: the values outside parenthesis are mean values, and those in parenthesis are standard errors.}
    \end{tabular}
\end{table*}

    A trained rater read each campaign's web page for evidence to answer the 26 RWW questions and follow the criteria to give ratings. Table \ref{tab:RWWAvgRating} presents the statistics of the RWW ratings based on our campaign samples. The level of evidence detail may be determined by the extent of development but also the preference of the creators to share such details. It is reasonable that creators might selectively disclose some aspects of their products and projects with more details and disguise others. For instance, for questions 2, 17, 22, and 26, the average ratings are below 0.03, indicating that not enough information can be found for these factors in the campaign descriptions. The ratings to these four questions are therefore removed during our later statistical analysis. Table \ref{tab:ModelControlVar} reports the descriptive statistics of the control variables.

\begin{table*}[h!]
    \centering
    \caption{Summary statistics for control variables.}
    \label{tab:ModelControlVar}
    \begin{tabular}{clllllll}
    \hline
    \textbf{Variables} &
      \multicolumn{1}{c}{\textbf{\begin{tabular}[c]{@{}c@{}}3D printers \\ on Indiegogo\end{tabular}}} &
      \multicolumn{1}{c}{\textbf{\begin{tabular}[c]{@{}c@{}}3D printers \\ on Kickstarter\end{tabular}}} &
      \multicolumn{1}{c}{\textbf{\begin{tabular}[c]{@{}c@{}}Smart watches \\ on Indiegogo\end{tabular}}} &
      \multicolumn{1}{c}{\textbf{\begin{tabular}[c]{@{}c@{}}Smart watches \\ on Kickstarter\end{tabular}}} &
      \multicolumn{1}{c}{\textbf{\begin{tabular}[c]{@{}c@{}}All 3D \\ printers\end{tabular}}} &
      \multicolumn{1}{c}{\textbf{\begin{tabular}[c]{@{}c@{}}All smart \\ watches\end{tabular}}} &
      \multicolumn{1}{c}{\textbf{All}} \\ \hline
    \textbf{\begin{tabular}[c]{@{}c@{}}Funding \\ Raised\end{tabular}} &
      \begin{tabular}[c]{@{}l@{}}37,957.81\\ (16,194.62)\end{tabular} &
      \begin{tabular}[c]{@{}l@{}}344,679.89\\ (94,821.62)\end{tabular} &
      \begin{tabular}[c]{@{}l@{}}117,776.19\\ (65,641.55)\end{tabular} &
      \begin{tabular}[c]{@{}l@{}}676,108.04\\ (440,848.71)\end{tabular} &
      \begin{tabular}[c]{@{}l@{}}222,777.53\\ (59,747.62)\end{tabular} &
      \begin{tabular}[c]{@{}l@{}}379,850.33\\ (211,233.26)\end{tabular} &
      \begin{tabular}[c]{@{}l@{}}283,380.42\\ (89,131.83)\end{tabular} \\ \hline
    \textbf{\begin{tabular}[c]{@{}c@{}}Funding \\ Goal\end{tabular}} &
      \begin{tabular}[c]{@{}l@{}}67,650.19\\ (33,331.21)\end{tabular} &
      \begin{tabular}[c]{@{}l@{}}70,196.64\\ (12,560.61)\end{tabular} &
      \begin{tabular}[c]{@{}l@{}}131,477.85\\ (30,600.81)\end{tabular} &
      \begin{tabular}[c]{@{}l@{}}84,676.13\\ (12,757.22)\end{tabular} &
      \begin{tabular}[c]{@{}l@{}}69,184.59\\ (15,127.49)\end{tabular} &
      \begin{tabular}[c]{@{}l@{}}109,509.69\\ (17,468.97)\end{tabular} &
      \begin{tabular}[c]{@{}l@{}}84,743.09\\ (11,567.29)\end{tabular} \\ \hline
    \textbf{\begin{tabular}[c]{@{}c@{}}Funded \\ Percent\end{tabular}} &
      \begin{tabular}[c]{@{}l@{}}1.70\\ (0.46)\end{tabular} &
      \begin{tabular}[c]{@{}l@{}}7.33\\ (1.90)\end{tabular} &
      \begin{tabular}[c]{@{}l@{}}5.09\\ (1.19)\end{tabular} &
      \begin{tabular}[c]{@{}l@{}}1.36\\ (0.84)\end{tabular} &
      \begin{tabular}[c]{@{}l@{}}7.14\\ (4.39)\end{tabular} &
      \begin{tabular}[c]{@{}l@{}}4.07\\ (2.12)\end{tabular} &
      \begin{tabular}[c]{@{}l@{}}4.70\\ (1.10)\end{tabular} \\ \hline
    \textbf{Characters} &
      \begin{tabular}[c]{@{}l@{}}9,270.71\\ (1,300.15)\end{tabular} &
      \begin{tabular}[c]{@{}l@{}}10,688.57\\ (1,025.14)\end{tabular} &
      \begin{tabular}[c]{@{}l@{}}9,178.44\\ (1,260.87)\end{tabular} &
      \begin{tabular}[c]{@{}l@{}}11,515.87\\ (917.60)\end{tabular} &
      \begin{tabular}[c]{@{}l@{}}10,125.06\\ (803.93)\end{tabular} &
      \begin{tabular}[c]{@{}l@{}}10,298.46\\ (800.43)\end{tabular} &
      \begin{tabular}[c]{@{}l@{}}10,191.12\\ (581.62)\end{tabular} \\ \hline
    \textbf{Figures} &
      \begin{tabular}[c]{@{}l@{}}6.68\\ (0.98)\end{tabular} &
      \begin{tabular}[c]{@{}l@{}}12.66\\ (1.19)\end{tabular} &
      \begin{tabular}[c]{@{}l@{}}17.81\\ (2.72)\end{tabular} &
      \begin{tabular}[c]{@{}l@{}}19.17\\ (1.87)\end{tabular} &
      \begin{tabular}[c]{@{}l@{}}10.28\\ (0.88)\end{tabular} &
      \begin{tabular}[c]{@{}l@{}}18.45\\ (1.68)\end{tabular} &
      \begin{tabular}[c]{@{}l@{}}13.43\\ (0.91)\end{tabular} \\ \hline
    \textbf{Tables} &
      \begin{tabular}[c]{@{}l@{}}0.52\\ (0.20)\end{tabular} &
      \begin{tabular}[c]{@{}l@{}}1.09\\ (0.24)\end{tabular} &
      \begin{tabular}[c]{@{}l@{}}1.19\\ (0.34)\end{tabular} &
      \begin{tabular}[c]{@{}l@{}}0.35\\ (0.16)\end{tabular} &
      \begin{tabular}[c]{@{}l@{}}0.86\\ (0.17)\end{tabular} &
      \begin{tabular}[c]{@{}l@{}}0.80\\ (0.20)\end{tabular} &
      \begin{tabular}[c]{@{}l@{}}0.83\\ (0.13)\end{tabular} \\ \hline
    \textbf{Videos} &
      \begin{tabular}[c]{@{}l@{}}0.84\\ (0.10)\end{tabular} &
      \begin{tabular}[c]{@{}l@{}}2.34\\ (0.28)\end{tabular} &
      \begin{tabular}[c]{@{}l@{}}1.62\\ (0.50)\end{tabular} &
      \begin{tabular}[c]{@{}l@{}}1.78\\ (0.26)\end{tabular} &
      \begin{tabular}[c]{@{}l@{}}1.74\\ (0.19)\end{tabular} &
      \begin{tabular}[c]{@{}l@{}}1.69\\ (0.29)\end{tabular} &
      \begin{tabular}[c]{@{}l@{}}1.72\\ (0.16)\end{tabular} \\ \hline
    \textbf{Rewards} &
      \begin{tabular}[c]{@{}l@{}}7.94\\ (0.89)\end{tabular} &
      \begin{tabular}[c]{@{}l@{}}12.15\\ (0.97)\end{tabular} &
      \begin{tabular}[c]{@{}l@{}}9.19\\ (1.12)\end{tabular} &
      \begin{tabular}[c]{@{}l@{}}9.30\\ (0.87)\end{tabular} &
      \begin{tabular}[c]{@{}l@{}}10.47\\ (0.72)\end{tabular} &
      \begin{tabular}[c]{@{}l@{}}9.24\\ (0.71)\end{tabular} &
      \begin{tabular}[c]{@{}l@{}}10.00\\ (0.52)\end{tabular} \\ \hline
    \textbf{\begin{tabular}[c]{@{}c@{}}Team Intro \\ {[}Yes=1/No=0{]}\end{tabular}} &
      \begin{tabular}[c]{@{}l@{}}0.45\\ (0.09)\end{tabular} &
      \begin{tabular}[c]{@{}l@{}}0.38\\ (0.07)\end{tabular} &
      \begin{tabular}[c]{@{}l@{}}0.77\\ (0.08)\end{tabular} &
      \begin{tabular}[c]{@{}l@{}}0.59\\ (0.11)\end{tabular} &
      \begin{tabular}[c]{@{}l@{}}0.41\\ (0.06)\end{tabular} &
      \begin{tabular}[c]{@{}l@{}}0.69\\ (0.07)\end{tabular} &
      \begin{tabular}[c]{@{}l@{}}0.52\\ (0.04)\end{tabular} \\ \hline
    \textbf{\begin{tabular}[c]{@{}c@{}}Timeline \\ {[}Yes=1/No=0{]}\end{tabular}} &
      \begin{tabular}[c]{@{}l@{}}0.32\\ (0.09)\end{tabular} &
      \begin{tabular}[c]{@{}l@{}}0.43\\ (0.07)\end{tabular} &
      \begin{tabular}[c]{@{}l@{}}0.69\\ (0.09)\end{tabular} &
      \begin{tabular}[c]{@{}l@{}}0.86\\ (0.07)\end{tabular} &
      \begin{tabular}[c]{@{}l@{}}0.38\\ (0.06)\end{tabular} &
      \begin{tabular}[c]{@{}l@{}}0.77\\ (0.06)\end{tabular} &
      \begin{tabular}[c]{@{}l@{}}0.53\\ (0.04)\end{tabular} \\ \hline
    \multicolumn{8}{l}{Note: the values outside parenthesis are mean values, and those in parenthesis are standard errors.}
    \end{tabular}
\end{table*}

\subsection{Baseline Model}
    We first explored the critical factors and a baseline prediction model regardless of platform and product differences. By comparing the average RWW ratings of the campaigns of two platforms or two product categories using a t-test, the factors that exhibit non-significant differences across platforms or products are first identified. Then, we use these factors as candidate predictive variables, together with all the control variables, to train linear regression models that predict the amount of crowdfunding raised. By using K-fold cross-validation through step-wise screening (sifting the predictive variables while always keeping all control variables), we shortlisted the RWW factors that are most significantly associated with the amount of crowdfunding raised and at the same time derive the regression model with the highest prediction power, as presented in Table \ref{tab:ModelBaseline}. The model fits our sample data with an overall $R^2$ of 64\% and an adjusted $R^2$ of 58\%.

\begin{table}[h!]
    \centering
    \caption{The baseline model of crowdfunding prediction.}
    \label{tab:ModelBaseline}
    \begin{tabular}{ll}
    \hline
    \textbf{Variables}                    & \textbf{All Observations}       \\ \hline
    Intercept                             & 1.97 (0.49)                     \\
    \textbf{Category {[}3DP=0/SW=1{]}}    & \textbf{0.62 (0.01)}            \\
    \textbf{Platform {[}IGG=0/KS=1{]}}    & \textbf{-1.01 (\textless{}.01)} \\
    \# of figures                         & \textless{}0.1 (0.99)           \\
    \# of tables                          & 0.06 (0.63)                     \\
    \# of videos                          & -0.15 (0.15)                    \\
    \# of rewards                         & 0.05 (0.12)                     \\
    Team Intro {[}Yes=1/No=0{]}           & 0.17 (0.68)                     \\
    Timeline {[}Yes=1/No=0{]}             & 0.22 (0.63)                     \\
    ln(Goal)                              & 0.14 (0.40)                     \\
    ln(Chars)                             & 0.33 (0.25)                     \\ \hline
    \textbf{Voice of customer - Q01}      & \textbf{2.09 (0.01)}            \\
    \textbf{Functional feasibility - Q08} & \textbf{1.29 (0.01)}            \\
    \textbf{Value propositions - Q12}     & \textbf{1.91 (0.01)}            \\
    \textbf{Risk evaluation - Q16}        & \textbf{1.67 (0.04)}            \\
    \textbf{Growth alignment - Q25}       & \textbf{1.31 (0.04)}            \\ \hline
    Self-prediction $R^2$                 & 0.635                           \\
    Adjusted Self-prediction $R^2$        & 0.580                           \\ \hline
    \end{tabular}
\end{table}

\begin{figure}[h!]
    \centering
    \includegraphics[width=0.9\linewidth]{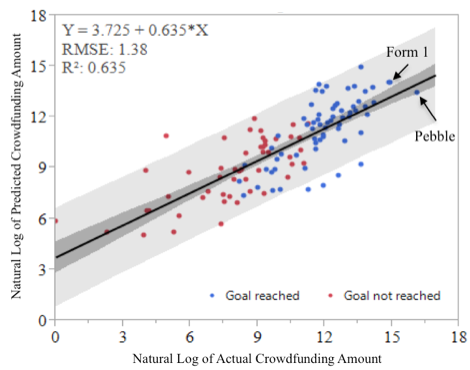}
    \caption{The predicted crowdfunding amount vs. the actual crowdfunding amount in the natural logarithm. Red dots represent failed campaigns.}
    \label{fig:ModelPlot}
\end{figure}

    Figure \ref{fig:ModelPlot} shows the regression line within the 90\% confidence intervals, which cover almost all samples in our data set. The two data points on the upper right most of the regression line are the Pebble smart watch and Form 1 3D printer. Figure \ref{fig:ModelPlot} also shows that our model can well predict the amounts of funding raised for campaigns that reached their goals and those that did not. Creators may apply the prediction model on their to-be-launched new campaigns to predict potential funding to be raised based on the intrinsic characteristics of their projects. If the predicted funding is much lower than the creators' expectations and needs, the creators are warned to further improve the project before launching the campaign. The predicted funding level may also guide creators to set reasonable and achievable funding goals. It may particularly help avoid situations in which a considerable amount of funding is raised with validated backer interests, but the creators still fail to collect funding because the goal was set too high. 

    Specifically, the baseline model includes five critical RWW factors to predict the amount of crowdfunding raised regardless of the platform and product differences. Their influences on the crowdfunding amount are statistically significant, as evidenced by the small p-values for their coefficients. 
    
    Q01 asks, ``Is there an adequate voice-of-customer type of evidence,'' indicating the importance of understanding customer needs for product innovation that can attract potential backers. Q08 asks, ``Is there adequate evidence of functional feasibility with available/breakthrough technology/material,'' implying the importance to convince the backers that the product functions can be achieved. These two factors are related to how \textbf{real} the product is in the eyes of the backers. 
    
    Q12 asks, ``Is there adequate tangible or intangible advantages offered to the customers,'' focusing on value propositions as the product must provide clear benefits to the backers. Q16 asks, ``Is there an adequate evaluation of the vulnerabilities of the product advantages,'' emphasizing the importance of risk evaluation in the eyes of the backers on crowdfunding. These two factors are related to how likely the product can eventually \textbf{win}. 
    
    Finally, Q25 asks, ``Is there adequate evidence that the product supports an overall growth strategy,'' which is about growth alignment, indicating that the new product needs to support and is driven by the longer-term growth strategy of the startup. The growth alignment factor is related to whether the product is \textbf{worth} developing.
    
    Therefore, these five critical factors (voice of customer, functional feasibility, value propositions, risk evaluation, and growth alignment) cover the Real, Win, and Worth categories, respectively. Note that, these RWW factors are critical to but not unique to the crowdfunding research, because the RWW framework, despite being new to crowdfunding research, is a collection of factors previously known from traditional contexts. To convince potential backers online, these critical factors deserve special attention. Creators should ensure project excellence at least for this subset of factors and provide sufficient evidence correspondingly on the campaign web page.

\subsection{Platform-specific and Product-specific Models}
    We further explore platform- or product-specific models and the corresponding critical factors, aiming to be more predictive in specific settings. For this purpose, we consider the five critical factors in the baseline model as control variables in each product- or platform-specific prediction model. The remaining RWW factors are analyzed as predicting factors and filtered through step-wise K-fold cross-validations using individual product-specific and platform-specific data samples. The resultant critical factors in the respective platform- and product-specific models are summarized in Table \ref{tab:ModelCritical}. When fitted with specific product or platform data, each of these specific models exhibits improved predictability than the baseline model, as measured in $R^2$ reported in the last two rows of Table \ref{tab:ModelCritical}.

\begin{table*}[h!]
    \centering
    \caption{Critical factors in the platform- and product-specific prediction models. Each model is trained on specific data samples. RWW factors are denoted with ``X'' signs where p-values are less than 0.05.}
    \label{tab:ModelCritical}
    \begin{tabular}{rccccc}
    \hline
    \multicolumn{1}{c}{\textbf{Estimates}} &
      \textbf{\begin{tabular}[c]{@{}c@{}}Baseline \\ Model\end{tabular}} &
      \textbf{\begin{tabular}[c]{@{}c@{}}Kickstarter \\ Model\end{tabular}} &
      \textbf{\begin{tabular}[c]{@{}c@{}}Indiegogo \\ Model\end{tabular}} &
      \textbf{\begin{tabular}[c]{@{}c@{}}3D Printer \\ Model\end{tabular}} &
      \textbf{\begin{tabular}[c]{@{}c@{}}Smart Watch \\ Model\end{tabular}} \\ \hline
    \multicolumn{1}{c}{\textbf{Baseline RWW Factors}} &       &       &       &       &       \\
    \textit{Voice of customer} - Q01                           & X     &       &       &       & X     \\
    \textit{Functional feasibility} - Q08                      & X     & X     & X     & X     &       \\
    \textit{Value propositions} - Q12                          & X     & X     &       & X     &       \\
    \textit{Risk evaluation} - Q16                             & X     & X     & X     &       &       \\
    \textit{Growth alignment} - Q25                            & X     &       &       &       &       \\ \hline
    \multicolumn{1}{c}{\textbf{Specific RWW Factors}} &       &       &       &       &       \\
    \textit{Market demography} - Q03                           &       & X     & X     &       &       \\
    \textit{Development compatibility} - Q07                   &       &       & X     &       &       \\
    \textit{Unique advantage} - Q13                            &       & X     &       &       &       \\
    \textit{Patent strategy} - Q14                             &       &       & X     &       &       \\
    \textit{Patent maintenance} - Q15                          &       &       & X     &       &       \\
    \textit{Enhanced perception} - Q18                         &       &       & X     &       &       \\ \hline
    Self-prediction $R^2$                             & 0.635 & 0.789 & 0.759 & 0.666 & 0.771 \\
    Adjusted Self-prediction $R^2$                    & 0.580 & 0.724 & 0.589 & 0.586 & 0.656 \\ \hline
    \end{tabular}
\end{table*}

    Different sets of additional RWW factors are found critical for a specific product category or crowdfunding platform. For Kickstarter, market demography (Q03) and unique advantage (Q13) are critical, whereas market demography (Q03), development compatibility (Q07), patent strategy (Q14), patent maintenance (Q15) and enhanced perception (Q18) are critical for Indiegogo. The resultant predictabilities of the Kickstarter-specific and Indiegogo-specific models are 0.789 and 0.759, respectively, which are much higher than the baseline model's predictability for Kickstarter (0.650) and Indiegogo (0.433). The addition of platform-specific RWW factors improved the predictability.
    
    More factors are critical to Indiegogo than to Kickstarter. Creators aiming to run campaigns on Indiegogo might need to demonstrate more evidence than those on Kickstarter in order to convince potential backers. For example, as shown in Table \ref{tab:ModelCritical}, enhanced perception (Q18) plays a critical role in raising crowdfunding on Indiegogo, but not the case for Kickstarter. This factor measures the effectiveness of a customer's understanding of the product's value proposition, which is standard on the creativity-focused Kickstarter platform. The differentiated critical factors may also indicate that the state of development of Kickstarter campaigns is higher than that of Indiegogo campaigns on average \cite{Henard2001WhyOthers}.
    
    As shown in the last two columns of Table \ref{tab:ModelCritical}, no additional product-specific RWW factors are found for 3D printers and smart watches. However, there are slight differences in the factor weights and statistical significance in these two prediction models. For 3D printers, functional feasibility (Q08) and value propositions (Q12) are strongly correlated with crowdfunding raised, indicating the importance of providing evidence that product innovation is technically feasible and offers definite value to the customers. Backers are likely to concern these factors because 3D printers are still new to the consumer market despite its wide industrial and laboratory applications. For smart watches, the voice-of-customer (Q01) becomes the single RWW factor with statistical significance. Evidence of customer opinions on a new smart watch is fundamentally essential in the eyes of backers.
    
    In general, the product- and platform-specific models present higher predictability than the baseline model in their similar product and platform contexts, indicated by the $R^2$ values in the last two rows of Table \ref{tab:ModelCritical}. Context-specific models trained on context-specific data can provide tailored guidance for creators working in different product domains (e.g., 3D printers or smart watches) to develop crowdfunding campaigns on specific CFPs (e.g., Kickstarter or Indiegogo).

\section{Discussion}
\label{S:5}
    We have introduced a data-driven methodology to simultaneously derive a prediction model and identify critical factors for crowdfunding success, based on the Real-Win-Worth framework and public data of online reward-based crowdfunding campaigns. We have also demonstrated the methodology in the empirical contexts of 3D printer and smart watch campaigns on Kickstarter.com and Indiegogo.com. The methodology presents several novel contributions to the studies of crowdfunding, with a focus on its relevance to design processes and innovation projects, and to the studies of design innovation, by providing guidance to crowdfunding campaigns as an approach for design thinking.
    
    First, we are the first to adopt the Real-Win-Worth framework to the context of crowdfunding campaigns for design innovation projects. RWW assessment of reward-based campaigns addresses our focus on innovation instead of finance. Previously RWW was mainly used by large companies to evaluate their internal innovation projects. Notably, we newly developed 26 guiding questions to address RWW factors, together with rating criteria, to make the RWW assessment framework more actionable for data-driven research and practice. On this basis, RWW factor ratings were used to train a crowdfunding prediction model with public data, which is also novel because previously the RWW framework had been mostly used in discrete and qualitative manners. The 26 RWW questions and corresponding rating criteria can also be applied to evaluating general early-stage design innovation projects in companies, startups, or ad hoc design teams, and not limited to crowdfunding campaigns.
    
    Second, our research used the product and project descriptions on the web pages of reward-based campaigns as the primary data source to predict crowdfunding success. This new data focus also addresses the fundamental relevance of our methodology to design innovation processes. By contrast, prior studies on what influences crowdfunding success only analyzed such exogenous factors as text length, figure, table and video counts, which are not directly related to the innovation project itself. To predict crowdfunding success, we treat such exogenous factors as control variables only and instead focus on RWW factors (evaluated based on the content of campaign descriptions) as predictors as they are directly related to design innovation.
    
    Third, our use of the continuous variable of crowdfunding amounts as the success measure is a novel contribution to the literature. Prior studies typically treated crowdfunding success as a binary variable - meeting the fixed funding goal or not. The specific amount of crowdfunding is more accurate to gauge the actual level of engagements of early adopters into the design process and their interests in the design concept, thus being more relevant and crucial to inform design. By contrast, crowdfunding success or failure based on a fixed funding goal might be more meaningful to the interest in financing, because the funding can only be collected when the funding goal is reached.
    
    Furthermore, we have applied the new methodology to 3D printer and smart watch campaign data on Kickstarter.com and Indiegogo.com. The empirical study shows case the predictability of the trained models and makes sense of the identified critical factors in the specific product or platform contexts. The derived prediction models and critical factors can be directly valuable for 3D printer and smart watch innovators considering crowdfunding campaigns. Mainly, our analysis shows that the models trained with data more specific to a product category or crowdfunding platform present higher predictability. Therefore, creators are suggested to use data in their domains or contexts of practice for implementing the data-driven methodology.

\section{Conclusion}
\label{S:6}
    This research is motivated by the growing adoption of reward-based crowdfunding campaigns by designers or innovation teams to discover and engage early adopters, validate demands, seek feedback and learning for innovative designs. Despite its relevance to design thinking and processes, crowdfunding has been under-studied in the design and innovation literature, in contrast to the existence of many studies of crowdfunding in the venture capital and business literature \cite{Sorenson2016ExpandCrowdfunding, Mollick2016DemocratizingCrowdfunding, Agrawal2016AreCrowdfunding}. Previously there was no method or tool to guide designers and innovators in developing a crowdfunding campaign for product innovation. Campaigns of innovative products are faced with a high degree of uncertainty and often fail to engage early adopters and thus are ineffective to inform design.
    
    In this paper, we have introduced a data-driven method that designers and innovators can use to enhance their crowdfunding campaigns as part of the design innovation processes. Specifically, they can use the method to identify factors that are most critical for crowdfunding campaign success in their context and require strategic attention and to derive a prediction model that can evaluate the crowdfunding potential of their innovation projects. Both the critical factors and the prediction model are useful to guide and inform crowdfunding campaigns for innovative products. The novelty and value of our work arise mainly from the choices of analytical lens (RWW), data (campaign description), and measures (actual funding amount) most directly related to the innovation projects for crowdfunding. In turn, our work contributes to the crowdfunding literature from the design innovation perspective, and to the design literature by supporting crowdfunding campaigns as an approach for design thinking and processes.
    
    A few limitations and areas for future work are noteworthy. First, the manual campaign rating process is slow and a bottleneck for training on a large dataset. Future research may employ machine learning and natural language processing techniques for faster and more efficient campaign ratings. Second, additional exogenous factors, such as social network marketing \cite{Etter2013LaunchCampaigns, Greenberg2013CrowdfundingFailure}, could also influence crowdfunding success. The inclusion of these factors as control variables may improve prediction accuracy, but this effort also relies on the collection of related data. 
    
    Moreover, our case study only explored two product categories for a demonstration purpose. Thus readers should not view the empirical results as permanent or universal. Product domains differ and evolve over time. Analysis across more product domains and longer time spans may lead to a fundamental understanding of the critical factors regardless of domains and the shifts of critical factors across domains. Tests in broader contexts may also shed light on limitations of the methodology and opportunities to refine it. Also, our 26 RWW questions and corresponding rating criteria can be applied to evaluating other early-stage design innovation projects in companies, startups, or ad hoc design teams than those for crowdfunding. We plan to expand the scope of the empirical analysis in future work. Nevertheless, the research opens many doors for future opportunities for data-driven research and practice at the intersection of crowdfunding, design innovation, and entrepreneurship.

\section*{Acknowledgement}
\label{S:7}
    This research is supported by a grant from the SUTD-MIT International Design Centre at the Singapore University of Technology and Design, and SUSTech-MIT Joint Centers for Mechanical Engineering Research and Education.

\ifCLASSOPTIONcaptionsoff
  \newpage
\fi
\Urlmuskip=0mu plus 1mu\relax
\bibliographystyle{IEEEtran}
\bibliography{myRef}
\end{document}